\documentclass[aps,prd,showpacs,preprintnumbers]{revtex4}
\usepackage{graphicx}
\input epsf
\def\fmslash{\@ifnextchar[{\fmsl@sh}{\fmsl@sh[0mu]}}
\def\fmsl@sh[#1]#2{%
  \mathchoice
    {\@fmsl@sh\displaystyle{#1}{#2}}%
    {\@fmsl@sh\textstyle{#1}{#2}}%
    {\@fmsl@sh\scriptstyle{#1}{#2}}%
    {\@fmsl@sh\scriptscriptstyle{#1}{#2}}}
\def\@fmsl@sh#1#2#3{\m@th\ooalign{$\hfil#1\mkern#2/\hfil$\crcr$#1#3$}}
\makeatother \voffset -3cm
\begin{document}

\begin{flushright}
\end{flushright}

\title{Choice of heavy baryon currents in QCD sum
rules}
\author{Dao-Wei Wang}
\affiliation{Department of Applied Physics, Nat'l University of Defense
Technology, Hunan 410073, China}
\author{Ming-Qiu Huang}
\affiliation{CCAST (World Laboratory) P.O. Box 8730, Beijing, 100080}
\affiliation{and Department of Applied Physics, Nat'l University of Defense
Technology, Hunan 410073, China}
\date{\today}
\begin{abstract}
In this paper we investigate the effects due to the mixing of two
interpolating currents for ground state baryons within the framework of Heavy
Quark Effective Theory using the QCD sum rule approach. Both two-point and
three-point sum rules, thus the mass, coupling constant and Isgur-Wise
function sum rules are considered. It is interesting to contrast those
results with each other. Based on the Isgur-Wise functions obtained in this
paper, we also analyze the effects of current mixing to $\Lambda$- type and
and $\Sigma$-type semi-leptonic decays
$\Lambda_b\rightarrow\Lambda_cl\bar\nu$,
$\Sigma_b\rightarrow\Sigma_c\ell\bar\nu$ and
$\Sigma_b\rightarrow\Sigma^*_c\ell\bar\nu$. Decay widths corresponding to
various mixing parameters are obtained and can be compared to the
experimental data.
\end{abstract}
\pacs{13.30.-a, 14.20.-c, 12.39.Hg, 11.55.Hx} \maketitle
\section{Introduction}
\label{sec1}

Strong interactions between quarks can be well described by QCD in the
standard model. Recently important progresses in the theoretical description
of hadrons containing a heavy quark have been achieved with the development
of the Heavy Quark Effective Theory (HQET) \cite{HQET,review,KKP}. Based on
the spin-flavor symmetry of QCD, exactly valid in the infinite heavy quark
mass limit, $m_Q\rightarrow\infty$, this framework provides a systematic
expansion of heavy hadron spectra and both the strong and weak transition
amplitudes in terms of the leading contribution, plus corrections decreasing
as powers of $1/m_Q$. HQET has been applied successfully to learn about the
properties of meson and baryon made of both heavy and light quarks.

Due to the asymptotic freedom of QCD, non-perturbative effect
plays an important role in the hadronic physics. Thus it is
inevitable to employ some non-perturbative technique in strong
interaction related problems. QCD sum rule \cite{svzsum}
formulated in the framework of HQET \cite{hqetsum} is a desirable
approach and proves to be predictive \cite{Braun99}. This method
allows one to relate hadronic observable to QCD parameter {\it
via} the operator product expansion (OPE) of the correlator. The
choice of the interpolating current for a state with given spin
and parity is the first step in the application of QCD sum rule
method. Principally, if a current is chosen within the framework
of HQET, QCD sum rule can be applied to many fields without
ambiguity and successfully. But the real situation is not so
simple. The main problem lies just on the choice of the
interpolating currents. On the heavy meson side, the current
interpolating a given spin and parity ground state is unique for
it constitutes of one heavy and one light quark. However, on the
heavy baryon side, the interpolating current is not
unique\cite{DHLL,GrYa,Ivanov}. For a given state there exist two
commonly adopted interpolating currents. Both bear the general
form as \cite{GrYa,Ioff}
\begin{equation}
\label{cr} j^{\,v}=\epsilon_{abc}(q^{T\,a}_1\,C\Gamma \,\tau
\,q^{b}_2)\Gamma^\prime h_{v}^{c},
\end{equation}
in which $C$ is the charge conjugation matrix, $\tau$ is the
flavor matrix which is antisymmetric for $\Lambda_Q$ baryon and
symmetric for $\Sigma_{Q}^{(*)}$ baryon, $\Gamma$ and
$\Gamma^\prime$ are some gamma matrices, and a, b, c denote the
color indices. One kind of current's $\Gamma$ and $\Gamma^\prime$
can be chosen covariantly as
\begin{equation}
\label{lq} \Gamma=\gamma_5\;,  \hspace{1cm} \Gamma^\prime=1\;,
\end{equation}
for $\Lambda_Q$ baryon,
\begin{eqnarray} \label{sq}
&&{}\Gamma=\gamma_\mu\;,\hspace{1cm} \Gamma^\prime=
(\gamma_\mu+v_\mu)\,\gamma_5\;,
\end{eqnarray}
for $\Sigma_Q$ baryon, and
\begin{eqnarray}
\label{sqs} \Gamma=\gamma_\nu\;,\hspace{0.5cm}
\Gamma^\prime=-g_{\mu \nu}+\frac{1}{3}\gamma_\mu\,\gamma_\nu -
\frac{1}{3}(\gamma_\mu\,v_\nu-\gamma_\nu\,v_\mu)+\frac{2}{3}v_\nu\,v_\mu\;,
\end{eqnarray}
for $\Sigma_{Q}^{*}$ baryon. Another kind of current can be obtained by
inserting a factor $\rlap/v$ behind the $\Gamma$ matrices defined by
equations (\ref{lq})-(\ref{sqs}). We denote them as $j^{v}_{1}$ and
$j^{v}_{\,2}$, respectively. In QCD sum rule applications those two currents
are usually used separately\cite{DHLL,GrYa,WHL,DHHL,HLLS,GKY,CDNP}.
Constituent quark type current which is the linear combination of the two
previously defined currents with the same coefficient can also be found in
application\cite{GKY2}. But generally speaking, the interpolating current
should be the linear combination of the two currents with arbitrary
coefficients. And also there have been many papers treating just the question
of the choice of baryon currents both in full QCD and in HQET sum
rules\cite{Chung,Bagan}. In this paper we adopt the general form
$j^v=a\,j^{v}_{\,1}+b\,j^{v}_{\,2}$, in which the coefficients $a$ and $b$
can be arbitrary real numbers, to investigate the effects of different choice
of baryon currents on physical observable.

The baryon coupling constants in HQET are defined through the
vacuum-to-baryon matrix element of the interpolating current as
follows
\begin{eqnarray}
\label{cc}
 \langle 0\mid j^v \mid
\Lambda(v)\rangle&=&F_{\Lambda}\,u\,,\nonumber\\
 \langle 0\mid j^v \mid
 \Sigma(v)\rangle&=&F_{\Sigma}\,u\,,\nonumber\\
 \langle 0\mid j^v \mid
\Sigma^*(v)\rangle&=&\frac{1}{\sqrt{3}}\,F_{\Sigma^*}\,u^{\alpha}\;,
\end{eqnarray}
where $u$ is the spinor and $u_\alpha$ is the Rarita-Schwinger
spinor in the HQET, respectively. The coupling constants
$F_{\Sigma}$ and $F_{\Sigma}^{*}$ are equivalent since $\Sigma_Q$
and $\Sigma_Q^*$ belong to the doublet with the same spin-parity
of the light degrees of freedom.

The remainder of this paper is organized as follows. In Sec.
\ref{sec2} we focus our emphasis on two-point and three-point
correlators and thus obtained sum rules for ground state baryons.
In Sec. \ref{ssec1} mass sum rules and in Sec. \ref{ssec2} sum
rules for Isgur-Wise functions are presented. Sec. \ref{sec3} is
devoted to numerical results and our conclusions.

\section{Two-point and three-point correlators}
\label{sec2}
\subsection{Two-point correlator and mass sum rule}
\label{ssec1}

Two-point correlators are T-product of interpolating currents saturated
between vacuum
\begin{equation}
\label{twop} i \int dx\,e^{ik\cdot x}\langle 0\mid T\{j^v(x)\bar
j^v(0)\}\mid 0\rangle=\Gamma'\frac{1+\rlap/v}{2}\bar\Gamma'Tr[\tau
\tau^+]\Pi (\omega),
\end{equation}
where $k$ is the residual momentum and $\omega=2v\cdot k$. The QCD
sum rule determination of baryon coupling constants can be
achieved by analyzing the two-point correlator. These diagonal
correlators of the single interpolating currents have been
obtained long ago by many authors and are of the same form for
both the interpolating currents $j_1^v$ and $j_2^v$. Non-diagonal
correlators have been analyzed in Ref.\cite{GrYa} in the leading
order in $\alpha_s$ and in next to leading order in $\alpha_s$ in
Ref.\cite{GKY}. In our previous works\cite{WHL,DHHL,HLLS} we
adopted the diagonal correlator as the starting point since for
the non-diagonal correlator there is no perturbative contribution
under the usual assumption of quark-hadron duality, let alone to
be the dominant part to the sum rules derived. Here we only have
one unique interpolating current and there is only one diagonal
correlator and no non-diagonal case to be analyzed. Our
theoretical result thus is the combination of the previously
called diagonal and non-diagonal results. It should be noted that
the non-diagonal one is merely treated as power corrections of OPE
in our choice of current.

In our calculations condensates with a dimension higher than $6$
are not included for lack of information and radiative corrections
are out of consideration contemporarily. In order to obtain an
estimate of the dimension $3$ nonlocal quark condensate we adopt
the gaussian ansatz $\langle\bar q(0)q(x)\,\rangle=\langle\bar
qq\rangle \mbox{exp}(m_{0}^{2}x^2/16)$. Relevant Feynman diagrams
are presented in Fig. 1. Then it is straightforward to obtain the
two-point sum rules:
\begin{eqnarray}
\label{2psumrule}
2\,F^{2}_{\Lambda}\;e^{-2\bar\Lambda_{\Lambda}/T}&=&(a^2+b^2)\left[\frac{3\,T^6}{2^4\,\pi^4}
\,\delta_5(\omega_c/T)+\frac{T^2}{2^6\,\pi^2}\,\langle\frac{\alpha_s}{\pi}G^2\rangle\
+\frac{1}{3}\,\langle \bar q
q\rangle^2\,e^{-m_{0}^{2}/2T^2}\right]\nonumber\\&-&\frac{a\,b}{\pi^2}\langle
\bar q q\rangle
[T^3\,e^{-m_{0}^{2}/4T^2}+T\,\frac{m_{0}^{2}}{8}],\nonumber\\
\frac{2}{3}
F^{2}_{\Sigma}\;e^{-2\bar\Lambda_{\Sigma}/T}&=&(a^2+b^2)\left[\frac{3\,T^6}{2^4\;\pi^4}\,\delta_5(\omega_c/T)
-\frac{T^2}{3\,2^6\,\pi^2}\,\langle\frac{\alpha_s}{\pi}G^2\rangle+\frac{1}{3}\,\langle
\bar q
q\rangle^2\,e^{-m_{0}^{2}/2T^2}\right]\nonumber\\&-&\frac{a\,b}{\pi^2}\langle
\bar q q\rangle
[T^3\,e^{-m_{0}^{2}/4T^2}-T\,\frac{m_{0}^{2}}{24}].
\end{eqnarray}
Our two-point sum rules do agree with results previously obtained
in Refs.\cite{GrYa,WHL} . The functions $\delta_n(\omega_c/T)$
arise from the continuum subtraction and are given by
\begin{equation}
   \delta_n(x) = \frac{1}{n!}\int\limits_0^x\!\mbox{d}t\,
   t^n e^{-t} = 1 - e^{-x} \sum_{k=0}^n \frac{x^k}{k!} \,.
\end{equation}
The second term of the last equation is assigned to the continuum
mode, which can be much larger than the ground state contribution
for the typical value of parameter $\mbox{T}$ if the dimension of
the spectral densities are very high.

\subsection{Three-point sum rules}\label{ssec2}
For a heavy-heavy velocity changing current ${\cal J}=\bar h\Gamma\bar h'$ we
can define a three-point correlator in which $J$ is inserted between two
interpolating currents as below
\begin{eqnarray}
i^2\int dx_1dx_2 \;e^{i (k\cdot x_1-k'\cdot x_2)}\langle 0\mid
T\{j^{\,v}(x_1){\cal J}(0)\bar j^{\,v'}(x_2)\}\mid 0\rangle
&=&\nonumber\\\Gamma'\frac{1+\rlap/v}{2}\Gamma\frac{1+\rlap/v'}{2}\bar\Gamma'Tr[\tau
\tau^+]\;T(\omega,\omega',y)\;,\label{ms}
\end{eqnarray}
where $T(\omega,\omega',y)$ is analytic function in the ``off-shell energies"
$\omega=2v\cdot k$ and $\omega'=2v'\cdot k'$ with discontinuities for
positive values of these variables. It furthermore depends on the velocity
transfer $y=v\cdot v'$, which is fixed at its physical region for the process
under consideration. In the heavy quark limit, the matrix element of current
${\cal J}$ can be parameterized by one or two scalar functions of $y$. Those
scalar functions are called Isgur-Wise functions\cite{IW} and can be defined
as
\begin{equation}
\langle \Lambda_Q|\bar h\Gamma h'|\Lambda_{Q'}\rangle=\xi(y)\,\bar
u\,\Gamma\,u',
\end{equation}
for $\Lambda$-type baryon, and
\begin{equation}
\langle \Sigma_Q|\bar h\Gamma
h'|\Sigma_{Q'}\rangle=[-\xi_1(y)g_{\mu\nu}+\xi_2(y)v'_\mu v_\nu]\,\bar
\Psi_\mu\Gamma \,\Psi'_\nu,
\end{equation}
for $\Sigma$-type baryon, in which $u$ is the Dirac spinor as defined in
(\ref{cc}) and $\Psi_\mu$ is the covariant representation of the spin $1/2$
doublets $\Psi_\mu=u_\mu+\frac{1}{\sqrt{3}}(v_\mu+\gamma_\mu)u$. Both
$\xi(y)$ and $\xi_1(y)$ are normalized to unity at the zero recoil $y=1$ due
to the heavy quark symmetry. However, one cannot invoke symmetry arguments to
predict the normalization at $y=1$ of $\xi_2(y)$.

Saturating the three-point correlator with complete set of baryon states, one
can divide it into two parts. One is the part of interest, the contribution
of the lowest-lying baryon states associated with the heavy-heavy currents,
as one having poles in both the variables $\omega$ and $\omega'$ at the value
$\omega=\omega'=2\bar\Lambda$. The other contribution to the correlator comes
from higher resonant states. For the little knowledge of this part of
contribution it is common to resort to the quark-hadron duality, which
insures that continuum contribution can be approximated by the integral of
the perturbative spectral density over a continuum threshold, to get a
predictive result.

On theoretical side the scalar function $T(\omega,\omega',y)$ can be
calculated using quark and gluon language with vacuum condensates. Dispersion
relation enables one to express the correlator in the form of integrals of
the double spectral density as
\begin{equation}
T(\omega,\omega',y)=\int_{0}^{\infty}\int_{0}^{\infty}
\frac{\mbox{d}s}{s-\omega}\frac{\mbox{d}s'}
{s'-\omega'}\,\rho(\mbox{s},\mbox{s}',y).
\end{equation}
With the redefinition of the integral
variables\cite{DHLL,HLLS,IWfun,Neubert92D}
\begin{eqnarray}
s_+&=&\frac{s+s'}{2},\nonumber\\
s_-&=&\left(\frac{y-1}{y+1}\right)^{1/2}\frac{s-s'}{2},
\end{eqnarray}
the integration domain becomes
\begin{equation}
\int_{0}^{\infty}ds\int_{0}^{\infty}ds'=2\left(\frac{y-1}{y+1}\right)^{1/2}
\int_{0}^{\infty}ds_+\int_{-s_+}^{s_+}ds_-.
\end{equation}
It is in variable $s_+$ that the commonly adopted quark-hadron
duality is assumed\cite{Neubert92D,BS}
\begin{equation}
res.=2\left(\frac{y-1}{y+1}\right)^{1/2}
\int_{\omega'_c}^{\infty}ds_+\int_{-s_+}^{s_+}ds_-\frac{\rho(\mbox{s},\mbox{s}',y)}
{(s-\omega)(s'-\omega')},
\end{equation}
and for simplicity we take $\omega'_c$ to be equal to the
two-point continuum threshold $\omega_c$: $\omega'_c=\omega_c$.

In our theoretical calculations of the three-point correlator only
condensates with dimension no more than 6 are included. Order
$1/m_Q$ power corrections and radiative corrections are not
included in present calculations, either. For their contributions
to the correlator only amount to several percents and do not
change the numerical result dramatically. Also the gaussian ansatz
for the nonlocal quark condensate is adopted. Feynman diagrams
related to the calculations of three-point correlator are shown in
Fig. 2.

Then following the standard procedure we resort to the Borel
transformation $B^{\omega}_\tau$,
$B^{\omega^{\prime}}_{\tau^{\prime}}$ to suppress the
contributions of the excited states. Considered the symmetry of
the correlation function it is natural to set the parameters
$\tau$, $\tau^{\prime}$ to be the same and equal to $\mbox{2T}$,
where $\mbox{T}$ is the Borel parameter of the two-point sum
rules. Finally, we obtain the sum rules for the Isgur-Wise
functions as
\begin{eqnarray}
\label{iwfunc}
4F_{\Lambda}^{2}\,\xi(y)\,e^{-2\bar\Lambda_\Lambda/T}&=&\frac{3\,(a^2+b^2)}{\pi^4\,(1+y)^3}
T^6\delta_5(\omega_c/T)+
\frac{[a^2(2y+1)+b^2y(y+2)]\,T^2}{24(1+y)^2\pi^2}\langle\frac{\alpha_s}{\pi}G^2\rangle\nonumber\\&
+&\frac{2\,(a^2+b^2y)}{3} \langle\bar
qq\rangle^2e^{-(1+y)m_{0}^{2}/4T^2}-\frac{4a\,b\langle\bar q
q\rangle}{(1+y)\pi^2}
[T^3\,e^{-m_{0}^{2}(1+y)/8T^2}+T\,\frac{m_{0}^{2}}{24}(y+2)], \nonumber\\
\frac{4}{3}F_{\Sigma}^{2}\,\xi_1(y)\,e^{-2\bar\Lambda_\Sigma/T}&=&\frac{3\,(a^2+b^2)}
{\pi^4\,(1+y)^3}T^6\delta_5(\omega_c/T)
-\frac{(a^2+b^2)\,T^2}{24(1+y)^2\pi^2}\langle\frac{\alpha_s}{\pi}G^2\rangle\nonumber\\&
+&\frac{2\,(a^2+b^2y)}{3} \langle\bar
qq\rangle^2e^{-(1+y)m_{0}^{2}/4T^2} -\frac{4a\,b\langle \bar q
q\rangle}{(1+y)\pi^2}
[T^3\,e^{-m_{0}^{2}(1+y)/8T^2}-T\,\frac{m_{0}^{2}}{24}y], \nonumber\\
\frac{4}{3}F_{\Sigma}^{2}\,\xi_2(y)\,e^{-2\bar\Lambda_\Sigma/T}&=&\frac{3\,(a^2+b^2y)}
{\pi^4\,(1+y)^4}T^6\delta_5(\omega_c/T)
+\frac{(a^2y-b^2)\,T^2}{24(1+y)^3\pi^2}\langle\frac{\alpha_s}{\pi}G^2\rangle\nonumber\\&
+&\frac{2\,b^2}{3} \langle\bar qq\rangle^2e^{-(1+y)m_{0}^{2}/4T^2}
-\frac{4a\,b\langle \bar q q\rangle}{(1+y)^2\pi^2}
[T^3\,e^{-m_{0}^{2}(1+y)/8T^2}+T\,\frac{m_{0}^{2}}{24}(y-2)],
\end{eqnarray}
The unitary normalization of flavor matrix $Tr[\tau\tau^+]=1$ has been
applied to get those sum rules. It is obvious that the normalization of the
Isgur-Wise functions $\xi(y)$ and $\xi_1(y)$ at zero recoil is satisfied
automatically.

\section{numerical results and  conclusions}
\label{sec3}
It is obvious that in the expressions of two-point
and three-point sum rules the relative value of the two parameters
but not the absolute value plays an important role. So in this
section we change the current mixing parameters $a,b$ to one
angular variable $\theta$ through relations $b/a=\tan\theta$ where
$\theta$ can be restrained to the range $\theta\in[-\pi/2,\pi/2]$,
in which $\theta=0,\pm\pi/2$ correspond to the diagonal cases. In
the numerical analysis we will investigate the current mixing
effects in those sum rules. Standard values of the vacuum
condensates are
\begin{eqnarray}
&&\langle\bar q q\rangle = -(0.23\;\mbox{GeV})^3,\nonumber\\
&&\langle\frac{\alpha_s}{\pi}G^2\rangle = 0.012\;\mbox{GeV}^4,\nonumber\\
&&\langle\bar qg\sigma_{\mu\nu}G^{\mu\nu}q\rangle
=m_{0}^{2}\langle\bar q q\rangle,\;\;\;m_{0}^{2}=
0.8\;\mbox{GeV}^2.
\end{eqnarray}
They will be used in the following numerical analysis of the sum
rules.
\subsection{mass sum rules}
In the analysis of the coupling constant sum rules we need the
effective mass of the baryons in consideration as the input
parameter. One way to obtain this parameter is to extract it from
the experimental data assuming the heavy quark mass to be the
commonly recognized value from the outset of the analysis. For the
aim of consistency we adopt another way of obtaining the effective
mass parameter from Eq. (\ref{2psumrule}) which is based on the
QCD sum rule method entirely. The effective mass can be expressed
through the derivative of Borel variable T in the coupling
constant sum rules as
\begin{equation}
\bar\Lambda=-\frac{1}{2}\frac{\partial}{\partial
T^{-1}}\ln\,\mbox{K(T)},
\end{equation}
in which K(T) denotes the right hand side of Eq.(\ref{2psumrule}). So the
first step of our numerical analysis of those two-point sum rules is to find
the value of the effective mass. But the second step, which is the analysis
of coupling constant sum rules, will be omitted here as the focus of our
interest is on the mass sum rules entirely. Our main idea in the
consideration of the effect of the current mixing parameter to the sum rules
is to see if there exists a reasonable stability window of the Borel
parameter $\mbox{T}$ under the variation of $\theta$ in the range from
$-\pi/2$ to $\pi/2$. For the analysis of the coupling constant and mass sum
rules it is enough to take $\theta$ from $0$ to $\pi/2$ for on the range of
$\theta$ from $-\pi/2$ to $0$ the mass sum rules oscillate for the Borel
parameter $\mbox{T}\sim 1\;\mbox{GeV}$ so sharply that it is impossible to
find a desirable stability window. Thus we do not take into account of that
half part of $\theta$.

For the $\Lambda$-type baryon mass sum rule we find that there is no
agreeable stability window except around the vicinity of $\theta=0$ or
$\theta=\pi/2$, i.e. $a=1,b=0$ or $a=0,b=1$. So there is no or at most little
space left for the mixing of currents and what we obtained is the diagonal
sum rule. It is reasonable to assume this result does indicate that there
exists some mechanism which forbids the mixing of the two sector
interpolating currents in the mass sum rule in the leading order. The
diagonal sum rule result can be checked with previous work: When $\omega_c$
lies between $1.9\sim 2.5 \;\mbox{GeV}$ there exists the stability window
$0.35<\mbox{T}<0.65 \;\mbox{GeV}$. The effective mass thus obtained is
$\bar\Lambda_{\Lambda}=0.73\pm 0.07 \;\mbox{GeV}$, in which the error only
reflects the variation of Borel parameter $\mbox{T}$ and continuum threshold
$\omega_c$.

For the $\Sigma$-type baryon all sum rule windows are narrower than that of
$\Lambda$-type baryon and the stability is not as good as that of
$\Lambda$-type baryon, either\cite{GrYa}. With the increasing of $\theta$ the
stability falls drastically that the optional space left for the variation of
$\theta$ is smaller than that of $\Lambda$-type baryon. When $\omega_c$ lies
between $2.8\sim 3.3 \;\mbox{GeV}$ there exists the stability window
$0.4<\mbox{T}<0.7 \;\mbox{GeV}$ for the diagonal sum rules, which appear to
be the only surviving result with respect to the mixing of currents. The
effective mass thus obtained is $\bar\Lambda_\Sigma=0.90\pm 0.14
\;\mbox{GeV}$, in which the error only reflects the variation of Borel
parameter $\mbox{T}$ and continuum threshold $\omega_c$, too. Those results
can be checked with Refs.\cite{GrYa,WHL}. It is also worth noting that the
constituent quark type interpolating current cannot be distinguished from the
currents with arbitrary mixing parameters from the stability point of view.
Both $\Lambda$-type and $\Sigma$-type mass sum rules are presented in Fig. 3.
\subsection{sum rules for the three-point correlators}
\subsubsection{Isgur-Wise functions}
In order to get the numerical results for the Isgur-Wise
functions, we divide our three-point sum rules by two-point
functions to obtain $\xi$, $\xi_1$ and $\xi_2$ as functions of the
continuum threshold $\omega_c$ and the Borel parameter $\mbox{T}$.
This procedure can eliminate the systematic uncertainties and
cancel the dependence on mass parameter $\bar\Lambda$. As for the
mixing parameter $\theta$ in this part, we take it varying from
$-\pi/2$ to $\pi/2$ to determine the stability of the sum rules.

For the Isgur-Wise function of the $\Lambda$-type baryon,
$\xi(y)$, we find that it is not sensitive to the mixing parameter
$\theta$. Almost in the gamut of $\theta$ there exists a stability
window, and the stability does not change rapidly when $\theta$
goes far beyond the vicinity of the diagonal $\theta$s. The
continuum threshold is the same as that for the two-point sum rule
for the $\Lambda$-type baryon. For $\omega_c$ lies between
$1.9\sim 2.5 \;\mbox{GeV}$ there exists the stability window. The
stability window for $\xi(y=1.2)$ is a much narrower one,
$0.4<\mbox{T}<0.8 \;\mbox{GeV}$. The numerical results are shown
in Fig. 4. In the numerical analysis it is interesting that there
seems to exist a more stable window for the constituent quark type
current with $\theta=\pi/4$, though the tendency is not so
predominant.

The numerical analysis of the two Isgur-Wise functions of the
$\Sigma$-type baryon can be compared with each other. For the
function $\xi_1$, the existence of stability window can only allow
for the appearance of two diagonal sum rules and one constituent
quark type sum rule with mixing parameter $\theta=-\pi/4$. As for
the function $\xi_2$, the existence of stability window can allow
for the appearance of two diagonal sum rules besides one
constituent quark type sum rule with mixing parameter
$\theta=\pi/4$. When the continuum threshold $\omega_c$ lies
between $2.5\sim 3.3 \;\mbox{GeV}$ there exist the stability
windows for both functions with the allowed mixing angles. The
numerical results of two Isgur-Wise functions are shown in Fig. 5.
Due to lack of stability window of those two constituent quark
type sum rules with the mixing parameters $\theta=\pm\pi/4$ for
$\xi_1$ and $\xi_2$, both numerical results related to those two
sum rules are taken from the range $0.8<\mbox{T}<1.2$, the
continuum threshold is the same as that of the diagonal sum rules.
We also present our results for the function $\xi_2(1)$ in Table
~\ref{tab1}. Our results of the two constituent quark type
currents are approximately equal to 0.5 at the zero recoil, which
is consistent with the value obtained from constituent quark model
and large $N_c$ limit in Refs.\cite{korner,Chow}.
\begin{table}[htb]
\begin{tabular}{*{6}{@{\hspace{0.2cm}}c}}\hline\hline
$\theta$~~($\tan\theta=b/a$)&$\theta=0$&$\theta=
\pi/4$&$\theta=\pi/2$&$\theta=-\pi/4$&Refs.\cite{korner,Chow}\\\hline
$\xi_2(1)$&$0.40\pm0.07$&$0.51\pm0.01$&$0.62\pm0.07$&$0.48\pm0.01$&$0.5$\\\hline\hline
\end{tabular}
\caption{The value of Isgur-Wise function $\xi_2$ at the zero recoil.}
\label{tab1}
\end{table}

If we put the two Isgur-Wise functions which are normalized to unity at the
zero recoil into the linear form $\xi_{(1)}(y)=1-\rho_{(1)}^2(y-1)$, in which
the parameters $\rho^2$ and $\rho_1^2$ are the slopes (or charge radii) of
the Isgur-Wise functions, we can obtain the slopes $\rho^2$ and $\rho_1^2$
via a linear fit for $\xi(y)$ and $\xi_1(y)$ near the zero recoil. The final
results of the slopes are presented in Table ~\ref{tab2}. Many predictions
have been made on the value of the charge radii, and the results vary greatly
from each other\cite{Ivanov,DHHL,JMW,prd56-348,MNNFD,IL}.
\begin{table}[htb]
\begin{tabular}{*{5}{@{\hspace{0.5cm}}c}}\hline\hline
radii&$\theta=0$&$\theta=\pi/4$&$\theta=\pi/2$&$\theta=-\pi/4$\\\hline\hline
$\rho^2$&$0.66\pm0.08$&$0.46\pm0.03$&$0.35\pm0.13$&$0.41\pm0.10$\\
$\rho_1^2$&$0.80\pm0.07$&$0.58\pm0.08$&$0.57\pm0.18$&$0.30\pm0.11$\\\hline\hline
\end{tabular}
\caption{Charge radii $\rho^2$ and $\rho_1^2$ for the Isgur-Wise functions
$\xi(y)$ and $\xi_1(y)$.} \label{tab2}
\end{table}

\subsubsection{semi-leptonic decay rates}
With the appropriate forms for the Isgur-Wise functions $\xi,\, \xi_1\,
\mbox{and}\, \xi_2$ as we have derived in Eq. (\ref{iwfunc}), we can discuss
various semi-leptonic decays
involving the heavy-to-heavy transition $b\rightarrow c$. 
As some illustrative examples here we shall only consider three
types of semi-leptonic decays:
$\Lambda_b\rightarrow\Lambda_c\,\ell\bar\nu$,
$\Sigma_b\rightarrow\Sigma_c\,\ell\bar\nu$ and
$\Sigma_b\rightarrow\Sigma^*_c\,\ell\bar\nu$.

The semi-leptonic decay $\Lambda_b\rightarrow\Lambda_c\ell\bar\nu$
can be analyzed directly after obtaining the Isgur-Wise function
from the QCD sum rules. By neglecting the lepton mass, the
differential decay rate is \cite{DHHL}
\begin{eqnarray}
&&\frac{1}{\sqrt{y^2-1}}\frac{d\Gamma(\Lambda_b\rightarrow\Lambda_cl\bar\nu)}
{dy}=\frac{G_F^2|V_{cb}|^2m_{\Lambda_b}^{2}m_{\Lambda_c}^{3}}{(2\pi)^3}\nonumber\\
&&\times
\{(1-2ry+r^2)[(y-1)F_1^2+(y+1)G_1^2]+\frac{y^2-1}{3}(Ar^2+2Br+C)\},
\end{eqnarray}
where $r=m_{\Lambda_c}/m_{\Lambda_b}$. In the above equation,
\begin{eqnarray}
A&=&2F_1F_2+(y+1)F_2^2+2G_1G_2+(y-1)G_2^2,\nonumber\\
B&=&F_1^2+F_1F_2+F_2F_3+F_3F_1+yF_2F_3+G_1^2-G_1G_2-G_2G_3+G_3G_1+yG_2G_3,\nonumber\\
C&=&(y+1)F_3^2+2F_1F_3+(y-1)G_3^2-2G_1G_3.
\end{eqnarray}
To the next to leading order of $1/m_Q$, the form factors $F_i$
and $G_i$ bear the simple form
\begin{eqnarray}
F_1&=&C(\mu)\xi(y)+\left(\frac{\bar\Lambda}{2m_c}+\frac{\bar\Lambda}{2m_b}\right)[2\chi(y)+\xi(y)]
,\nonumber\\
G_1&=&C(\mu)\xi(y)+\left(\frac{\bar\Lambda}{2m_c}+\frac{\bar\Lambda}{2m_b}\right)[2\chi(y)
+\frac{y-1}{y+1}\xi(y)],\nonumber\\
F_2&=&G_2=-\frac{\bar\Lambda}{m_c(y+1)}\xi(y),\nonumber\\
F_3&=&-G_3=-\frac{\bar\Lambda}{m_b(y+1)}\xi(y),
\end{eqnarray}
where $C(\mu)$ is the perturbative QCD coefficient, $\chi(y)$ is
the sub-leading order Isgur-Wise function, which is only amounts
to the order of a few percents to the leading function and can be
safely neglected\cite{DHHL,KorMel}. With the form of the leading
order Isgur-Wise function (12), the differential decay rate of
$\Lambda_b\rightarrow\Lambda_cl\bar\nu$ is shown in Fig. 5. In
this analysis, we choose those related heavy quark masses to be:
$m_b=4.77\;\mbox{GeV},m_c=1.41\;\mbox{GeV}$\cite{WHL}, and
parameters $m_{\Lambda_b}=5.641\;\mbox{GeV},
m_{\Lambda_c}=2.285\;\mbox{GeV}$ can be found in Ref.\cite{data},
the renormalization point is $\mu=470 \;\mbox{MeV}$. It seems to
be inconsistent to use the quark masses obtained in Ref.\cite{WHL}
using factorization instead of gaussian ansatz to parameterize the
non-local quark condensate as done in Ref.\cite{DHLL} . But the
fact that the decay width is not sensitive to the heavy quark
masses allows us to use either pair of parameters without varying
the width significantly. The decay widths corresponding to four
typical mixing variables are listed in Table \ref{tab3}. Also
listed in Table \ref{tab3} are some predictions made by QCD sum
rule and other phenomenological approaches. The averages of the
decay widths are taken between $0.8<\mbox{T}<1.1 \;\mbox{GeV}$ and
$1.9\leq\omega_c\leq2.5 \;\mbox{GeV}$. Our results are in good
consistence with the experimental value,  $\Gamma=(4.0\pm
1.0)\times10^{-14}\mbox{GeV}$\cite{data}.

As to the two decays between $\Sigma$-type baryons, the decay widths have
simple and easy-to-be-interpreted forms when expressed with helicity
amplitudes. Related formulae can be found in many
references\cite{KKP,prd56-348,EIKL} and the decay widths corresponding to
various mixing parameters are listed in Table \ref{tab3}. In this part of
numerical analysis we only take into account of the contributions of the
leading order Isgur-Wise functions and omit the higher order effects. The
masses of the heavy baryons are taken to be $m_{\Sigma_b}=5.80
\;\mbox{GeV}$\cite{WHL}, $m_{\Sigma^*_c}=2.52 \;\mbox{GeV}$ and
$m_{\Sigma_c}=2.455 \;\mbox{GeV}$\cite{data}. For comparison we list the
results for those three types of decays predicted by QCD sum rule and other
phenomenological approaches in Table \ref{tab3}, too. It should be noted here
that function $\xi_1$ is the predominant part in the decay rates, so even
though for the mixing parameter $\theta=-\pi/4$ there exists no stability
window for function $\xi_2$, the total decay rates still have a mild
stability window for $\theta=-\pi/4$.
\begin{table}[htb]
\begin{tabular}{*{5}{@{\hspace{0.5cm}}c}}\hline\hline
\multicolumn{2}{c}{Refs.}&$\Lambda_b\rightarrow\Lambda_c\ell\bar\nu$
&$\Sigma_b\rightarrow\Sigma_c\ell\bar\nu$&$\Sigma_b\rightarrow\Sigma^*_c\ell\bar\nu$\\
\hline\hline
&$\theta=0$&$4.57\pm0.62$&$1.38\pm0.15$&$2.89\pm0.16$\\
&$\theta=\pi/4$&$3.98\pm0.07$&$1.66\pm0.11$&$2.78\pm0.06$\\
\raisebox{1.5ex}[0pt]{this paper}&$\theta=\pi/2$&$3.60\pm0.29$&$1.50\pm0.25$&$2.94\pm0.22$\\
&$\theta=-\pi/4$&$3.97\pm0.13$&$2.09\pm0.24$&$3.27\pm0.38$\\\hline\hline
\multicolumn{2}{c}{RTQM\,\,\cite{Ivanov}}&$4.08$&$1.51$&$2.48$\\
\multicolumn{2}{c}{RTQM\,\cite{prd56-348}}&$3.521$&$1.468$&$3.001$\\
\multicolumn{2}{c}{RTQM\,\,\cite{IL}}&$3.769$&$0.946$&$2.171$\\
\multicolumn{2}{c}{\hspace{0.2cm}QCM\cite{EIKL}{\,\,}}&$6.582$&$3.226$&\\
\multicolumn{2}{c}{SQM\,\,\cite{single}}&$3.883$&$2.830$&\\\hline\hline
\end{tabular}
\caption{Decay widths $\Gamma(\mbox{in}\, 10^{-14}\,\mbox{GeV})$ for the
semi-leptonic decays $\Lambda_b\rightarrow\Lambda_cl\bar\nu$,
$\Sigma_b\rightarrow\Sigma_cl\bar\nu$ and
$\Sigma_b\rightarrow\Sigma^*_cl\bar\nu$. Also presented in this table are
some phenomenological predictions using Relativistic Three-Quark Model(RTQM),
Quark Confinement Model(QCM) and Spectator Quark Model(SQM).} \label{tab3}
\end{table}

For conclusions, we have investigated the mixing of currents interpolating
ground heavy baryon state within the framework of HQET using QCD sum rule
approach. For the two-point sum rules there can only survive the diagonal
ones and the constituent quark type current is not preferable from the
stability point of view for both $\Lambda$-type and $\Sigma$-type baryons. As
for the three-point sum rules, Isgur-Wise function $\xi(y)$ for
$\Lambda$-type baryon is not sensitive to the mixing parameter and the
stability window exists almost for all the range of mixing parameter; on the
other hand, Isgur-Wise function $\xi_1(y)$ and $\xi_2(y)$ allows for two
diagonal, one constituent($\theta=\pi/4$) and one
anti-constituent($\theta=-\pi/4$) quark type sum rules. The effect of
different currents to semi-leptonic decays
$\Lambda_b\rightarrow\Lambda_cl\bar\nu$,
$\Sigma_b\rightarrow\Sigma_c\ell\bar\nu$ and
$\Sigma_b\rightarrow\Sigma^*_c\ell\bar\nu$ has also been analyzed in this
paper. We find that the current mixing effects in those processes are not
significant.
\begin{acknowledgments}
This work is supported in part by the National Natural Science Foundation of
China under Contract No. 19975068 and No. 10275091.
\end{acknowledgments}

\newpage
{\bf Figure Captions}
\begin{center}
\begin{minipage}{12cm}
{\sf Fig. 1.}{\quad  Non-vanishing diagrams for the two-point
correlator: (a) perturbative contribution, (b) quark-condensate,
(c) gluon-condensate, (d) mixed condensate, (e) four-quark
condensate contributions. The interpolating baryonic currents are
denoted by black circles. Heavy-quark propagators are drawn as
double curves.}
\end{minipage}
\end{center}

\begin{center}
\begin{minipage}{12cm}

{\sf Fig. 2.}{\quad Non-vanishing diagrams for the three-point
correlator: (a) perturbative contribution, (b) quark-condensate,
(c) and (d) gluon-condensate, (e) mixed condensate and (f)
four-quark condensate. The velocity-changing current operator is
denoted by a white square, and the interpolating baryonic currents
by black circles.}
\end{minipage}
\end{center}

\begin{center}
\begin{minipage}{120mm}
{\sf Fig. 3.}{\quad Sum rules for effective mass parameter
$\bar\Lambda$: The left one is for $\Lambda$-type baryon, in which
$\omega_c=2.2\;\mbox{GeV}$; and the right one is for $\Sigma$-type
baryon, in which $\omega_c=2.8\;\mbox{GeV}$.}
\end{minipage}
\end{center}

\begin{center}
\begin{minipage}{120mm}
{\sf Fig. 4.}{\quad Sum rules of the Isgur-Wise function $\xi(y)$
with various mixing parameters. The threshold in this figure is
$\omega_c=2.2\;\mbox{GeV} $ and the momentum transfer is $y=1.1$.}
\end{minipage}
\end{center}

\begin{center}
\begin{minipage}{12cm}
{\sf Fig. 5.}{\quad Sum rules of Isgur-Wise function $\xi_1(y)$
and $\xi_2(y)$ with various mixing parameters. The threshold in
this figure is $\omega_c=2.8\;\mbox{GeV}$ and the momentum
transfer is $y=1.1$.}
\end{minipage}
\end{center}

\begin{center}
\begin{minipage}{12cm}
{\sf Fig. 6.}{\quad Differential decay ratio of semi-leptonic
decay $\Lambda_b\rightarrow\Lambda_cl\bar\nu$ with various mixing
parameters as below: (a) $\theta=0$, (b) $\theta=\pi/4$, (c)
$\theta=\pi/2$ and (d)  $\theta=-\pi/4$. The solid, dashed and
dotted curves correspond to the threshold $\omega_c= 1.9,2.2,
2.5\;\mbox{GeV}$, respectively. And the Borel parameter is
$\mbox{T}=0.85\,\mbox{GeV}$ in this figure.}
\end{minipage}
\end{center}

\newpage
\begin{figure}
\epsfxsize=8cm \centerline{\epsffile{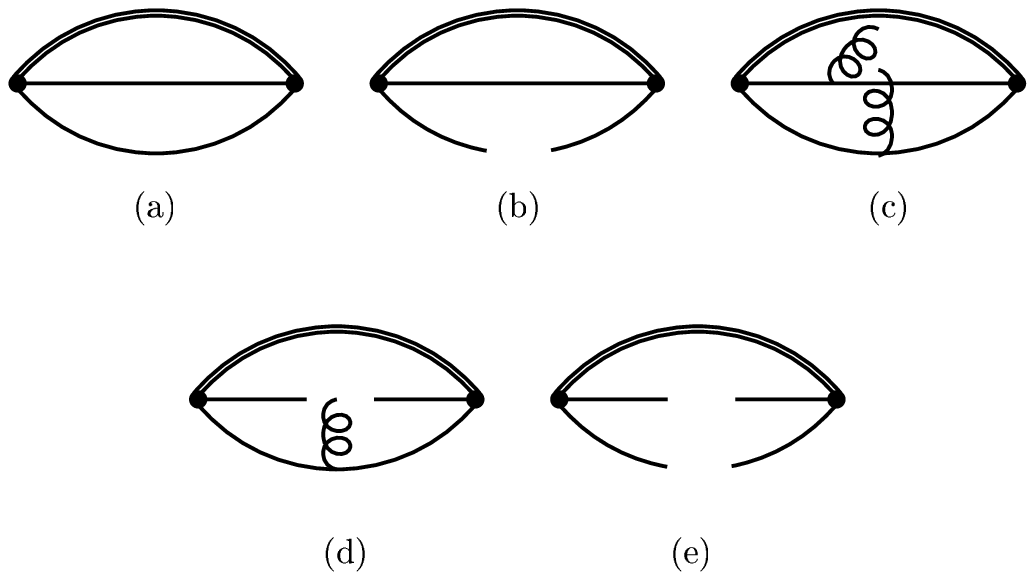}} \vspace{0.5cm}\caption{}
\end{figure}

\begin{figure}
\epsfxsize=8cm \centerline{\epsffile{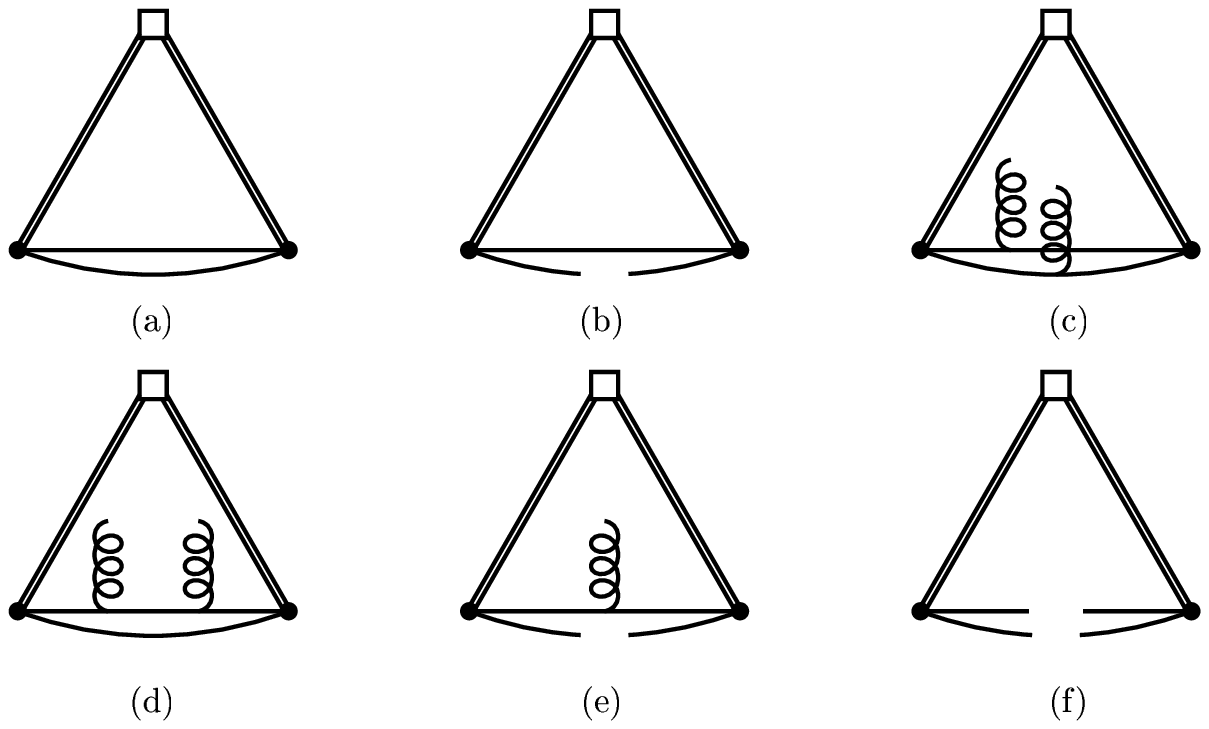}}\vspace{0.5cm} \caption{}
\end{figure}

\begin{figure}
\begin{minipage}{7cm}
\epsfxsize=7cm \centerline{\epsffile{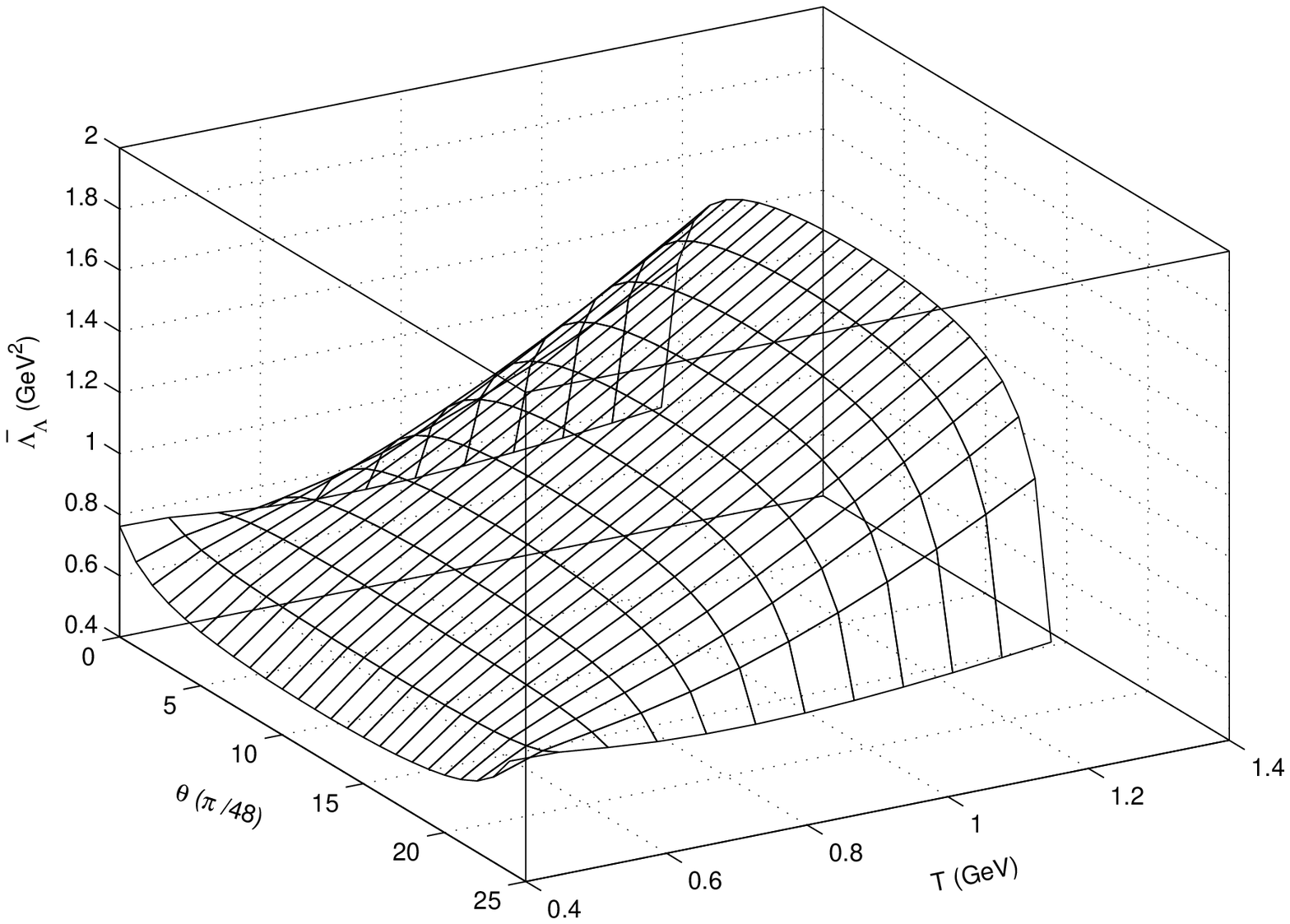}}
\end{minipage}
\begin{minipage}{7cm}
\epsfxsize=7cm \centerline{\epsffile{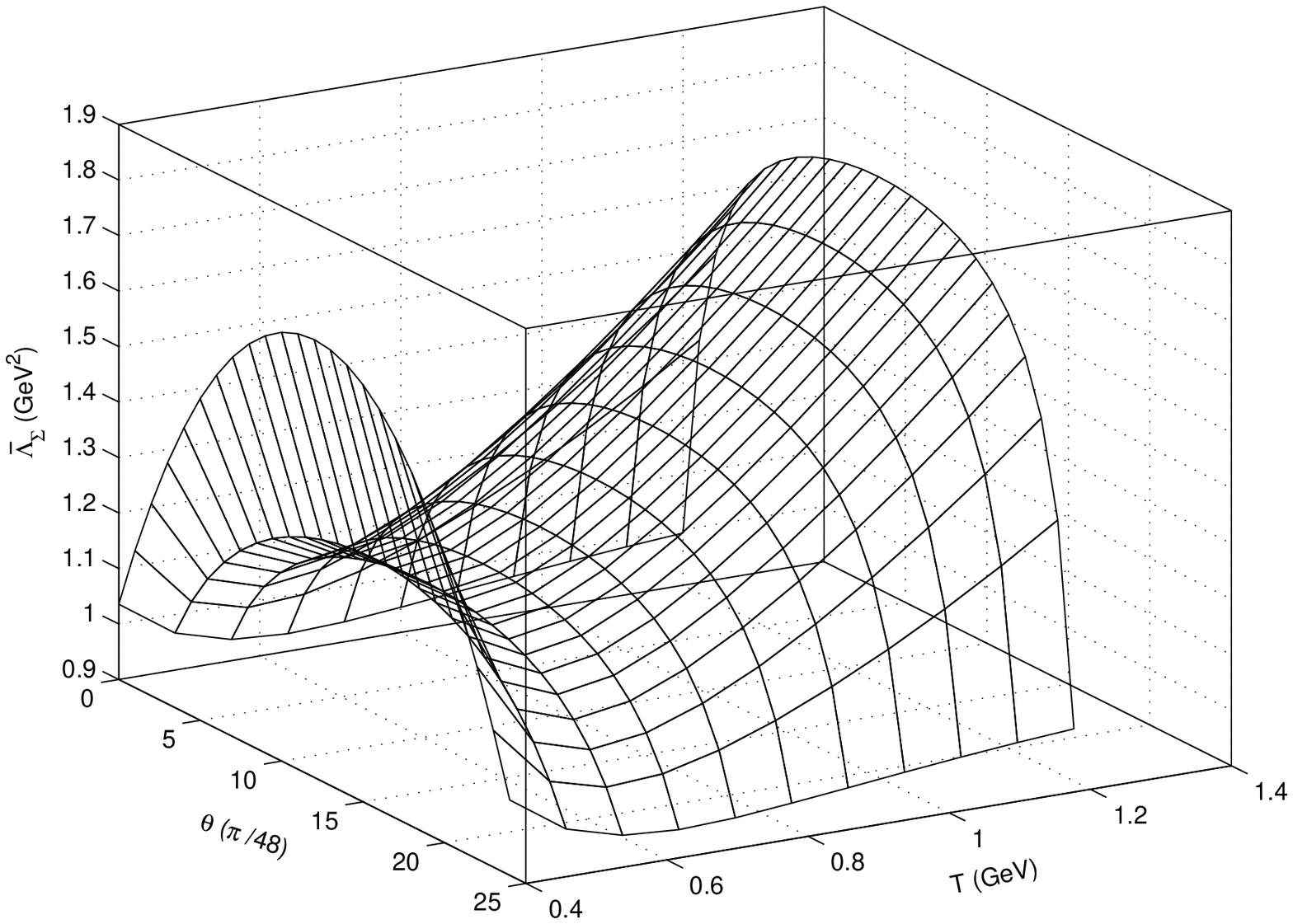}}
\end{minipage}
\caption{}
\end{figure}

\begin{figure}
\begin{minipage}{15cm}
\epsfxsize=7cm\hfill \centerline{\epsffile{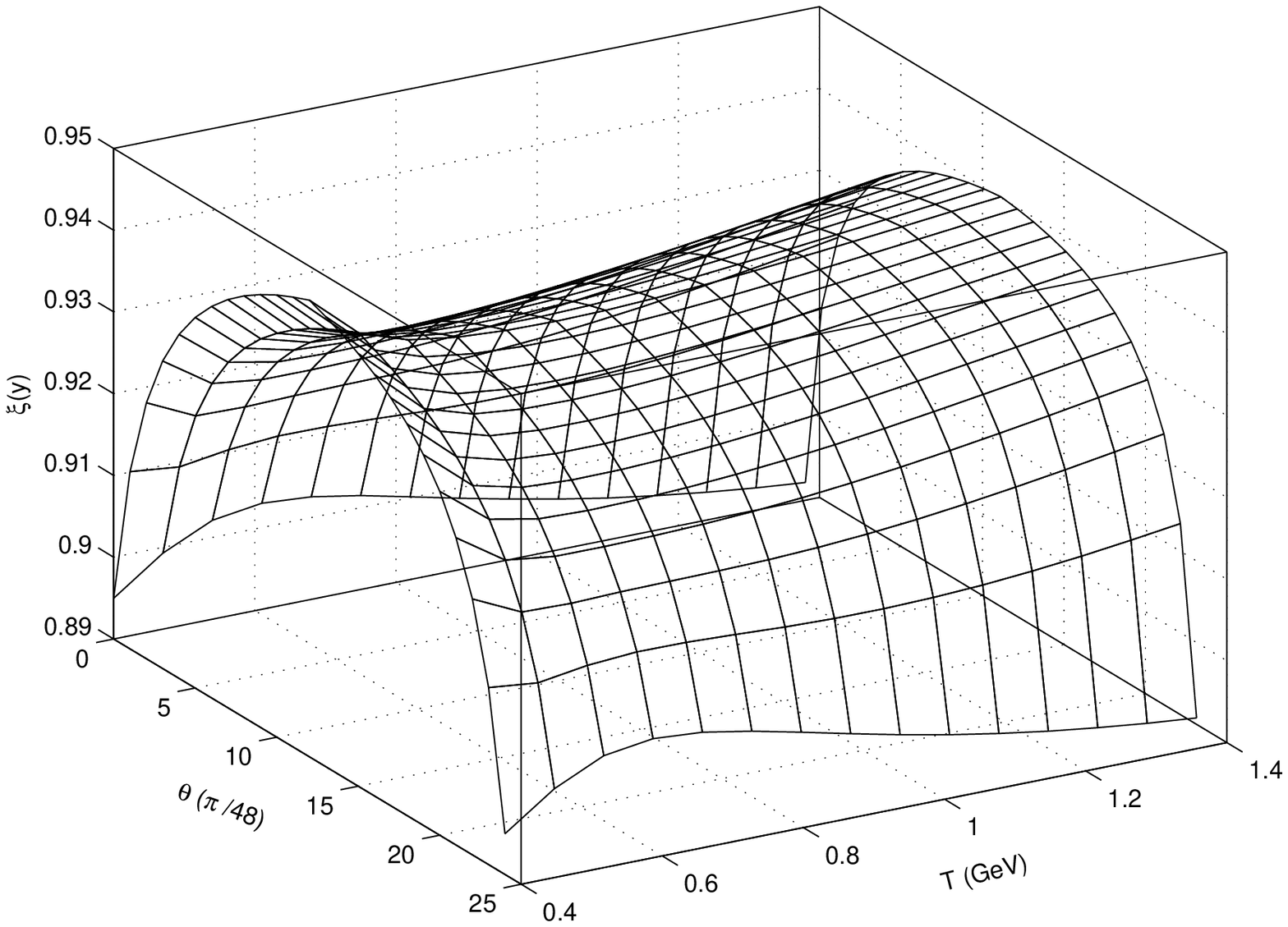}}
\end{minipage}
\caption{}
\end{figure}

\begin{figure}
\begin{minipage}{15cm}
\begin{minipage}{7cm}
\epsfxsize=7cm \centerline{\epsffile{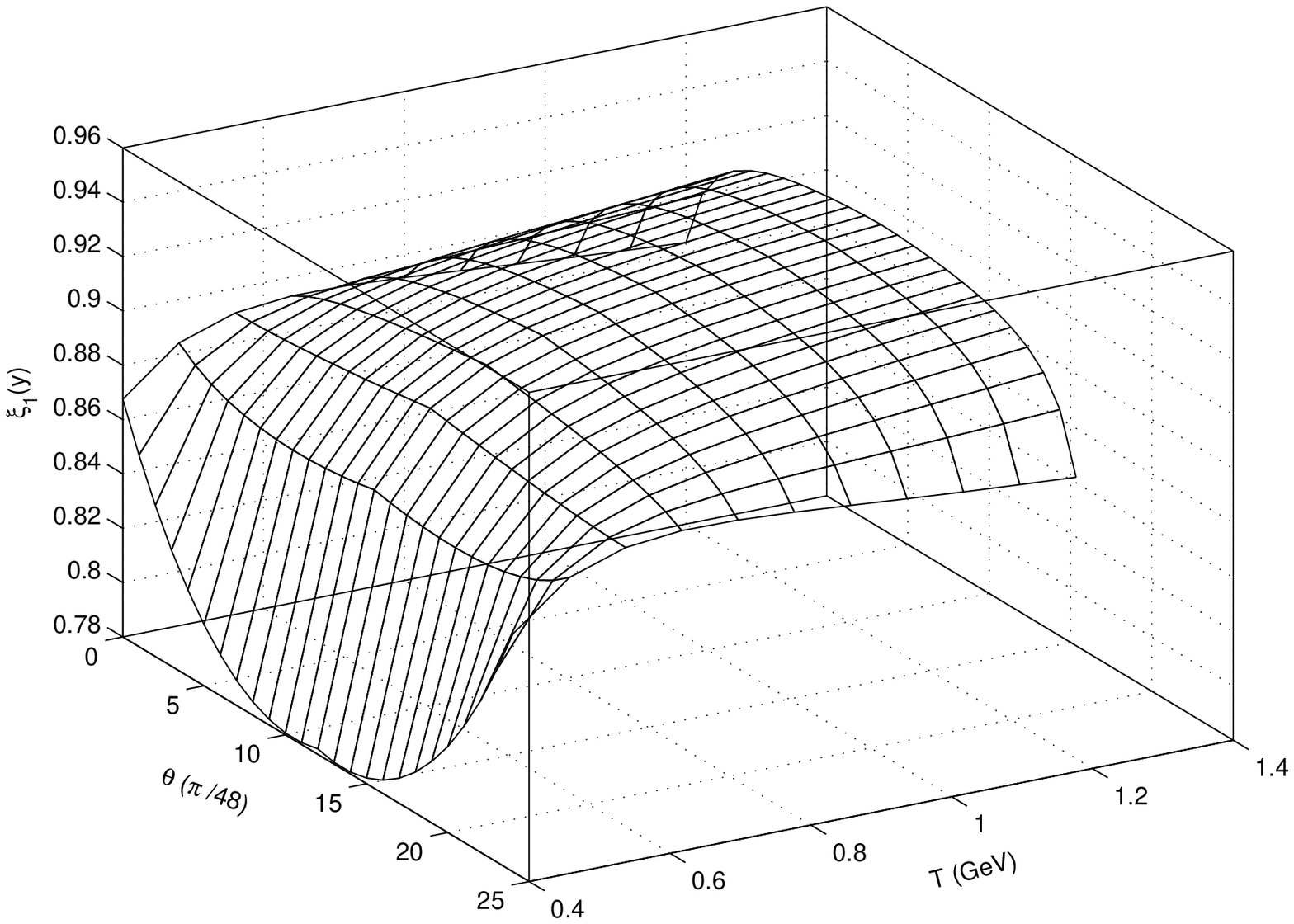}}
\end{minipage}
\begin{minipage}{7cm}
\epsfxsize=7cm \centerline{\epsffile{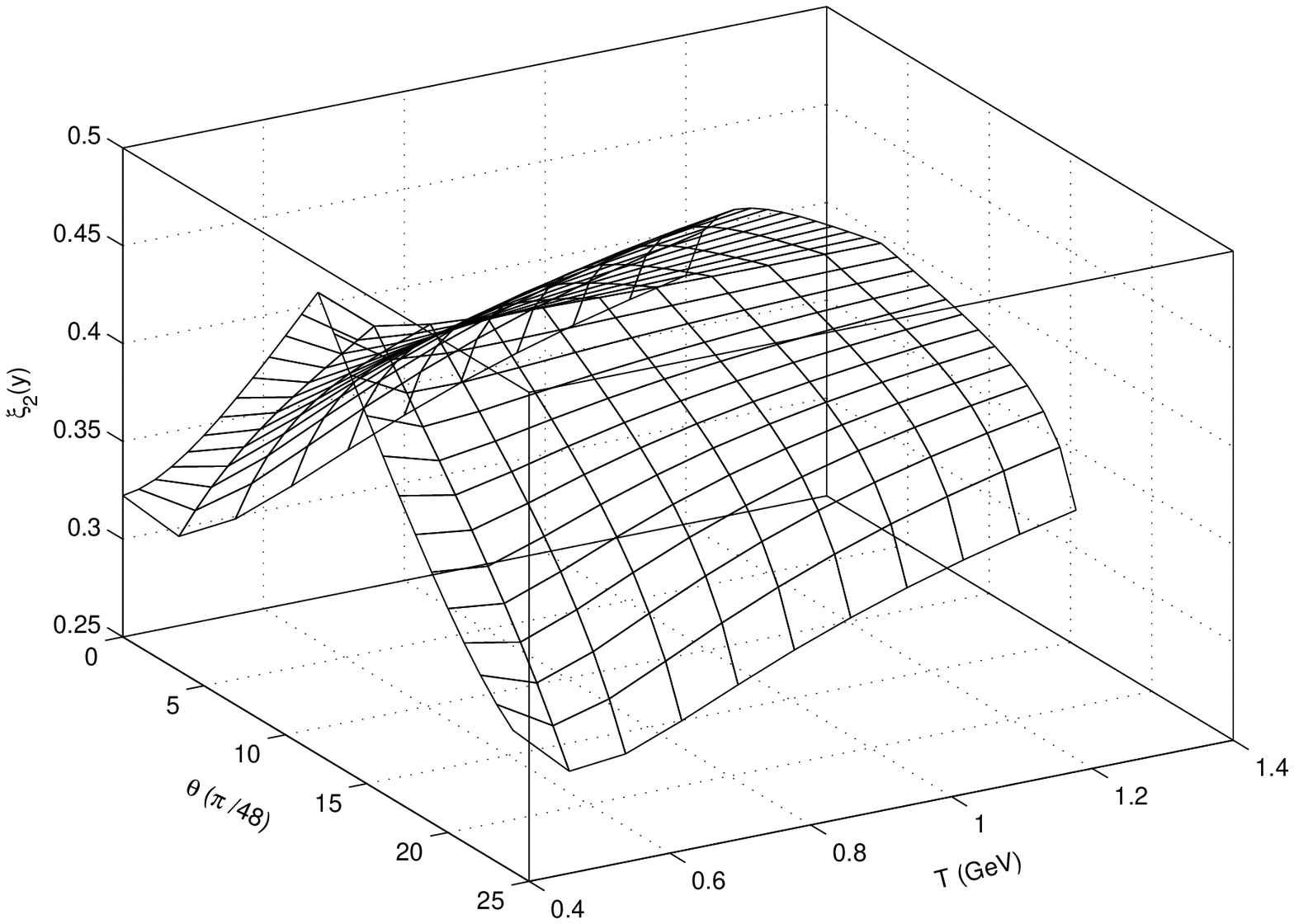}}
\end{minipage}
\end{minipage}
\caption{}
\end{figure}

\begin{figure}
\begin{minipage}{15cm}
\begin{minipage}{7cm}
\epsfxsize=7cm \centerline{\epsffile{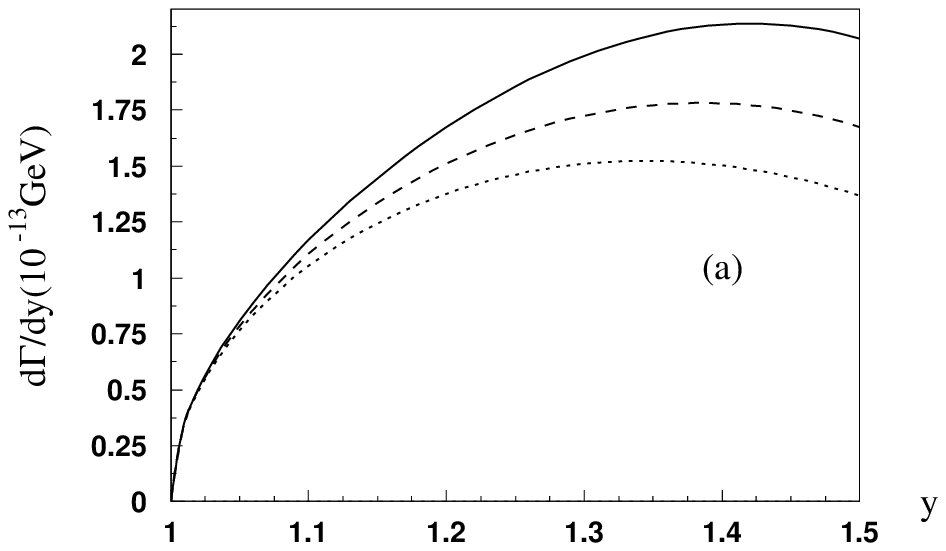}}
\end{minipage}
\begin{minipage}{7cm}
\epsfxsize=7cm \centerline{\epsffile{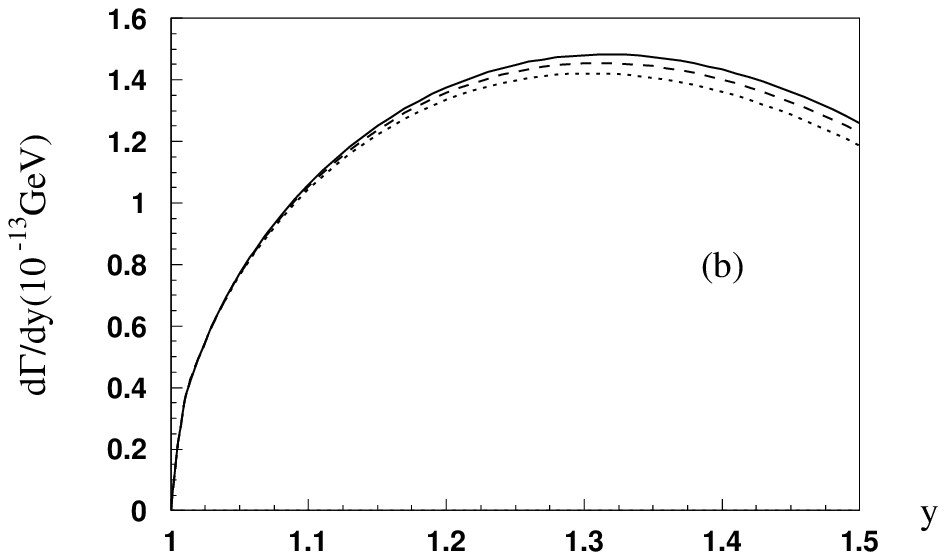}}
\end{minipage}
\end{minipage}
\begin{minipage}{15cm}
\begin{minipage}{7cm}
\epsfxsize=7cm \centerline{\epsffile{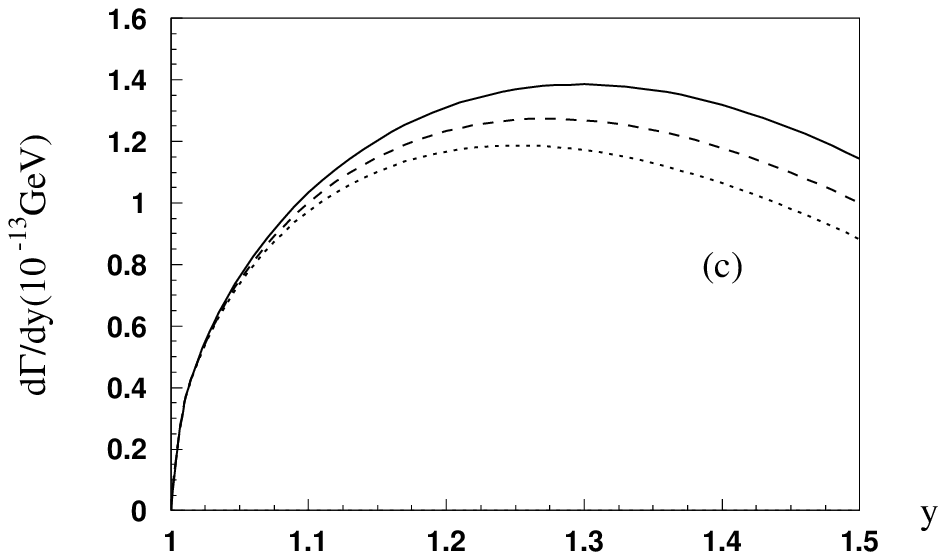}}
\end{minipage}
\begin{minipage}{7cm}
\epsfxsize=7cm \centerline{\epsffile{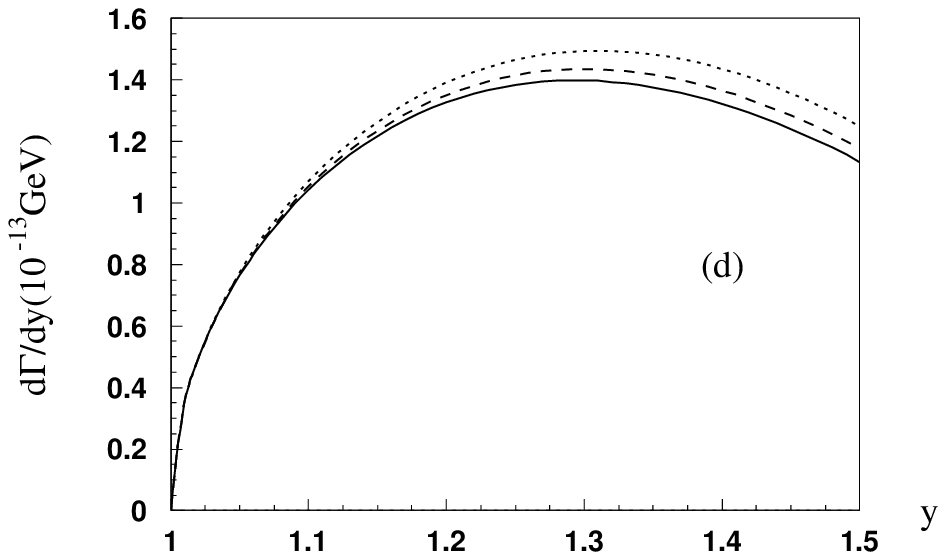}}
\end{minipage}
\end{minipage}
\caption{}
\end{figure}

\end{document}